\documentclass[sigconf]{acmart}

\usepackage{subcaption}
\usepackage[font=small]{caption}

\AtBeginDocument{\providecommand\BibTeX{{\normalfont B\kern-0.5em{\scshape i\kern-0.25em b}\kern-0.8em\TeX}}}

\setcopyright{acmcopyright}
\copyrightyear{2023}
\acmYear{2023}
\acmDOI{XXXXXXX.XXXXXXX}

\acmPrice{15.00}
\acmISBN{978-1-4503-XXXX-X/18/06}

\begin{document}

\title{Opportunities of Renewable Energy Powered DNN Inference}

\author{Seyed Morteza Nabavinejad}
\affiliation{\institution{Worcester Polytechnic Institute}
  \city{Worcester}
  \country{MA, USA}}
\email{snabavinejad@wpi.edu}

\author{Tian Guo}
\affiliation{\institution{Worcester Polytechnic Institute}
  \city{Worcester}
  \country{MA, USA}}
\email{tian@wpi.edu}

\newcommand{\tian}[1]{{\color{red}(Tian: #1)}}
\newcommand{\para }[1]{\emph {\textbf {#1}}}
\newcommand{\1}{{\em (i)}}
\newcommand{\2}{{\em (ii)}}
\newcommand{\3}{{\em (iii)}}
\newcommand{\4}{{\em (iv)}}
\newcommand{\5}{{\em (v)}}

\begin{abstract}
With the proliferation of the adoption of renewable energy in powering data centers, addressing the challenges of such energy sources has attracted researchers from academia and industry. One of the challenging characteristics of data centers with renewable energy is the intrinsic power fluctuation. Fluctuation in renewable power supply inevitably requires adjusting applications' power consumption, which can lead to undesirable performance degradation. This paper investigates the possible control knobs to manage the power and performance of a popular cloud workload, i.e., deep neural network inference, under the fluctuating power supply. Through empirical profiling and trace-driven simulations, we observe the different impact levels associated with inference control knobs on throughput, under varying power supplies. Based on our observations, we provide a list of future research directions to leverage the control knobs to achieve high throughput. 
\end{abstract}

\maketitle

\section{Introduction}
\label{sec:intro}

With the proliferation of renewable energy usage, many warehouse-scale infrastructures and cloud data centers have started to employ them as the main power source to manage carbon emissions. Renewable energy can help reduce the carbon footprint of data centers significantly. However, employing it to power data centers' computing, communication, and storage resources can be challenging due to its inherent power supply fluctuation. Fluctuating power generated by renewable sources affects the power capacity available to the underlying hardware and, consequently, the maximum power consumption of the applications deployed on those infrastructures. Therefore, there is a need to design power-aware mechanisms that can handle renewable energy fluctuation.

Deployment of AI-powered services, empowered by deep neural networks (DNNs), on warehouse-scale infrastructures and the cloud, is increasing. Therefore, embedding power-aware mechanisms in those services can help address the power fluctuation problem. Providers of such services seek \emph{high throughput} to serve more requests in a unit of time. They also desire high resource utilization to reduce operational costs and improve revenue. To this end, various hardware accelerators such as ASICs \cite{chen2016eyeriss, jouppi2017datacenter}, FPGA-based accelerators \cite{zhang2020dnnexplorer, roorda2022fpga}, and GPU-based accelerators \cite{li2022miso, crankshaw2017clipper, nabavinejad2022coordinated, choi2022serving} were proposed for improving DNN inference performance. In particular, GPUs have shown superior throughput for DNN inference and are widely used in today's data centers for serving inference requests. 
While these accelerators can significantly increase the throughput, they also consume a large share of power capacity. Therefore, managing their power consumption is key to effectively utilizing renewable energy.

To gain high throughput when accelerating DNN inference, a widely used approach is \emph{batching} \cite{nabavinejad2022coordinated, crankshaw2017clipper} where input data is processed in the form of batches instead of being processed one by one. Batching helps reuse the DNN model's parameters for several inputs and masks the overhead of copying input data to GPU memory \cite{shen2017escher, crankshaw2017clipper}. Batching has been recently used to manage the power consumption of GPU accelerators as well \cite{nabavinejad2022coordinated, yao2022eais}. Another popular alternative is \emph{multi-tenancy} \cite{xu2022igniter, gujarati2020serving,lemay2020perseus}, where several different DNNs are co-located on a single GPU. Multi-tenancy improves the throughput by sharing the computing resources between co-located DNNs. Previous works often focus on using multi-tenancy to improve GPU utilization, e.g., by squeezing instances of different DNNs to the same GPU~\cite{Dhakal2020-bl}. In this work, we explore multi-tenancy as a means to increase the per-model inference throughput. 
Both batching and multi-tenancy improve throughput by increasing resource utilization. However, this improved throughput, as expected, is achieved at the expense of increased power consumption. This paper \emph{explores the opportunities and trade-offs in using these software control knobs for DNN inference under power fluctuation}.

We first study the fluctuation of renewable power sources to emphasize the need for designing new algorithms and policies for managing renewable energy-powered DNN inference. 
Through empirical profiling on the NVIDIA RTX 3090 GPU using the MobileNet-V1 model, we observe the different impacts of controlling batching and multi-tenancy \emph{individually} and controlling them \emph{simultaneously} on power consumption and inference throughput. To further understand these knobs' potential under power fluctuations, we conduct trace-driven simulations with renewable power traces and empirically profiled data with ten CNNs. The results show that properly controlling the combination of batching and multi-tenancy achieves up to 86\% higher inference throughput than controlling batching or multi-tenancy individually.

Based on our observations, we outline avenues for future research. One direction is designing sampling-based approaches that can find a solution by only selecting and testing a few configurations instead of relying on complete profiling of all the possible configurations. Another possible direction is designing ML-based approaches to predict different configurations' throughput and power consumption and find a solution accordingly. For example, the performance of each configuration can be estimated based on its own features and the characteristics of the DNN model, such as computational complexity or architecture. We also emphasize the importance of reducing the power cap violation and suggest employing proactive or reactive approaches to address it.

\section{Motivation and Background}
\label{sec:motiv}

Low carbon emission policies and the high cost of fossil fuels have been driving the usage of renewable energies in warehouse-scale computers and data centers. While these energy sources can reduce operational costs, they bring challenges that need to be addressed. One big concern when using renewable energy is the natural fluctuation in the amount of power generated by renewable sources. The systems powered by renewable sources, such as large-scale DNN inference systems, should be able to adapt themselves to this fluctuation. There should be mechanisms to control the power consumption of such systems while maximizing their performance. In this section, first, we discuss the fluctuation in the amount of power generated by renewable energy sources. After that, we introduce different approaches to maximizing the throughput of DNN inference systems and their impact on the power consumption of underlying hardware platforms.

\subsection{Renewable Energy: Power Fluctuation}
\label{subsec:renew}

\begin{figure}[t]
    \centering
    \includegraphics[width=0.9\linewidth]{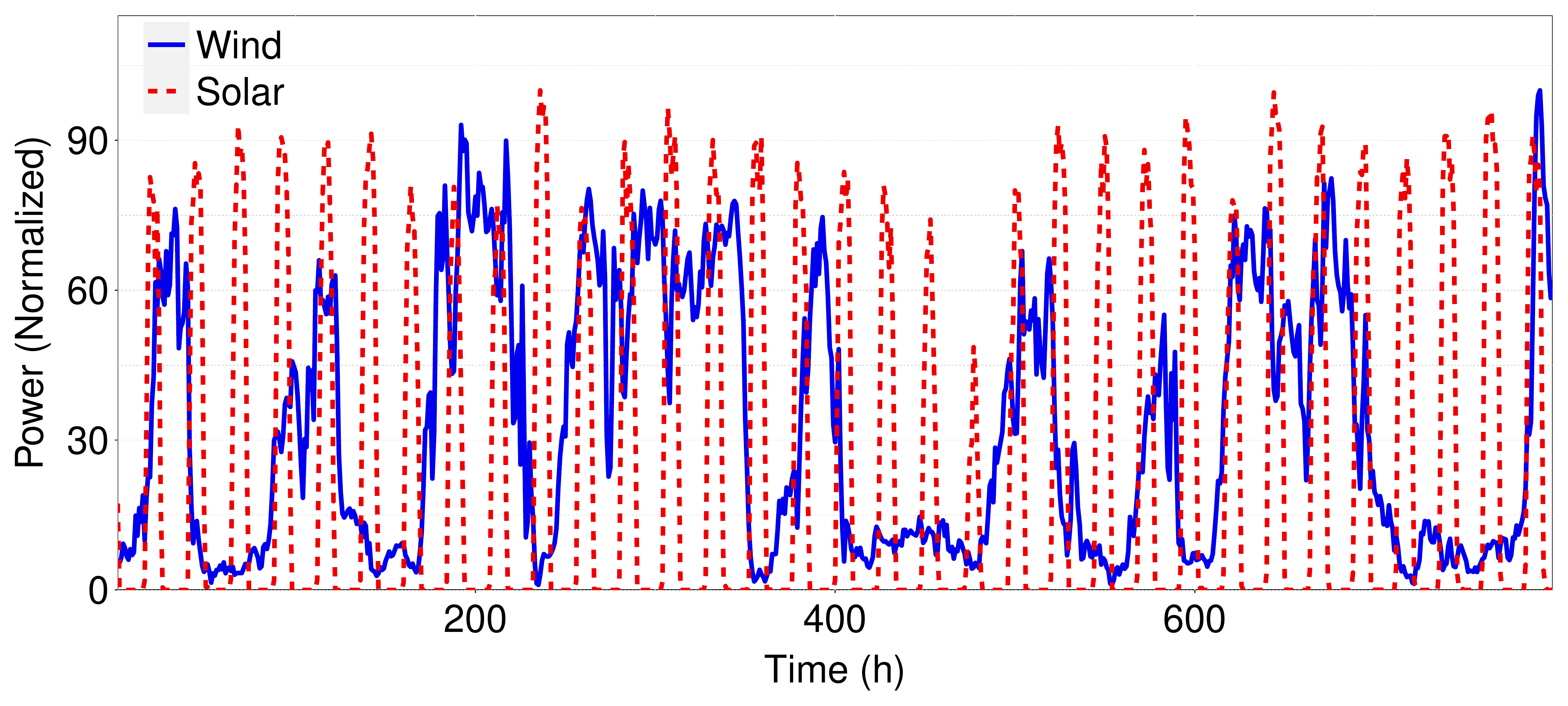}
\caption{Fluctuation in power generated by renewable sources.}
    \label{fig:motivPowerTrace}
\end{figure}

The amount of power generation capacity of renewable sources can significantly fluctuate at different scales: from minutes to days. To show this significant fluctuation, we use the \emph{US-CAL-CISO-2020} trace from Electricity Map \cite{carbonTrace} that provides the amount of power generated by different sources of renewable and non-renewable energy over the course of time in the California region during the year 2020. The trace contains the hourly average power for various power sources such as wind, solar, hydroelectric, geothermal, biomass, gas, and oil. We selected two renewable sources (wind and solar) and normalized the power values to between 0 and 100 to better visualize the fluctuations, as shown in Figure~\ref{fig:motivPowerTrace}. We can see a significant variation in the amount of power generated by both sources. Furthermore, we see a huge gap between the minimum and maximum power generated over the course of time. Also, the pattern of the amount of power generated and its fluctuation varies between solar and wind, indicating that they heavily depend on the power source. This challenging fluctuation needs to be addressed when designing different systems, including DNN inference serving systems, that are powered fully or partially by renewable energy. The systems should be able to adapt to the fluctuating power capacity, usually expressed as a power cap. Violating this power cap can impose different kinds of costs on the system, from monetary costs to increased carbon footprint.

\subsection{DNN Inference Control Knobs}
\label{subsec:BSMTL}

New generations of GPU accelerators used for DNN inference offer high computing and memory capacity. To leverage this capacity, two main techniques (batching and multi-tenancy) are employed by DNN inference applications. In batching, the DNN inference application is fed by batches of input data instead of each individual input separately (e.g., each image in image classification DNNs). Previous works have widely employed this approach \cite{crankshaw2017clipper, nabavinejad2022coordinated, gupta2020deeprecsys, yao2022eais} to increase the throughput and control the power/energy consumption or latency. With the capabilities of cutting-edge GPUs to support temporal and/or spatial co-location, it is possible to deploy several DNN models on the same GPU to leverage parallelism and achieve higher resource utilization and throughput. Multi-tenancy or co-location of several workloads or kernels on a single GPU and related challenges have been studied in a large body of research \cite{xu2022igniter, choi2022serving, li2022miso}. Another possible approach is combining batching and multi-tenancy, where several instances of the same DNN model are deployed on GPU with batch sizes greater than one and process the inference requests simultaneously. 

\begin{figure}[t]
\centering
    \subfloat[Throughput - Multi-Tenancy]
    {\includegraphics[width=0.45\linewidth]{{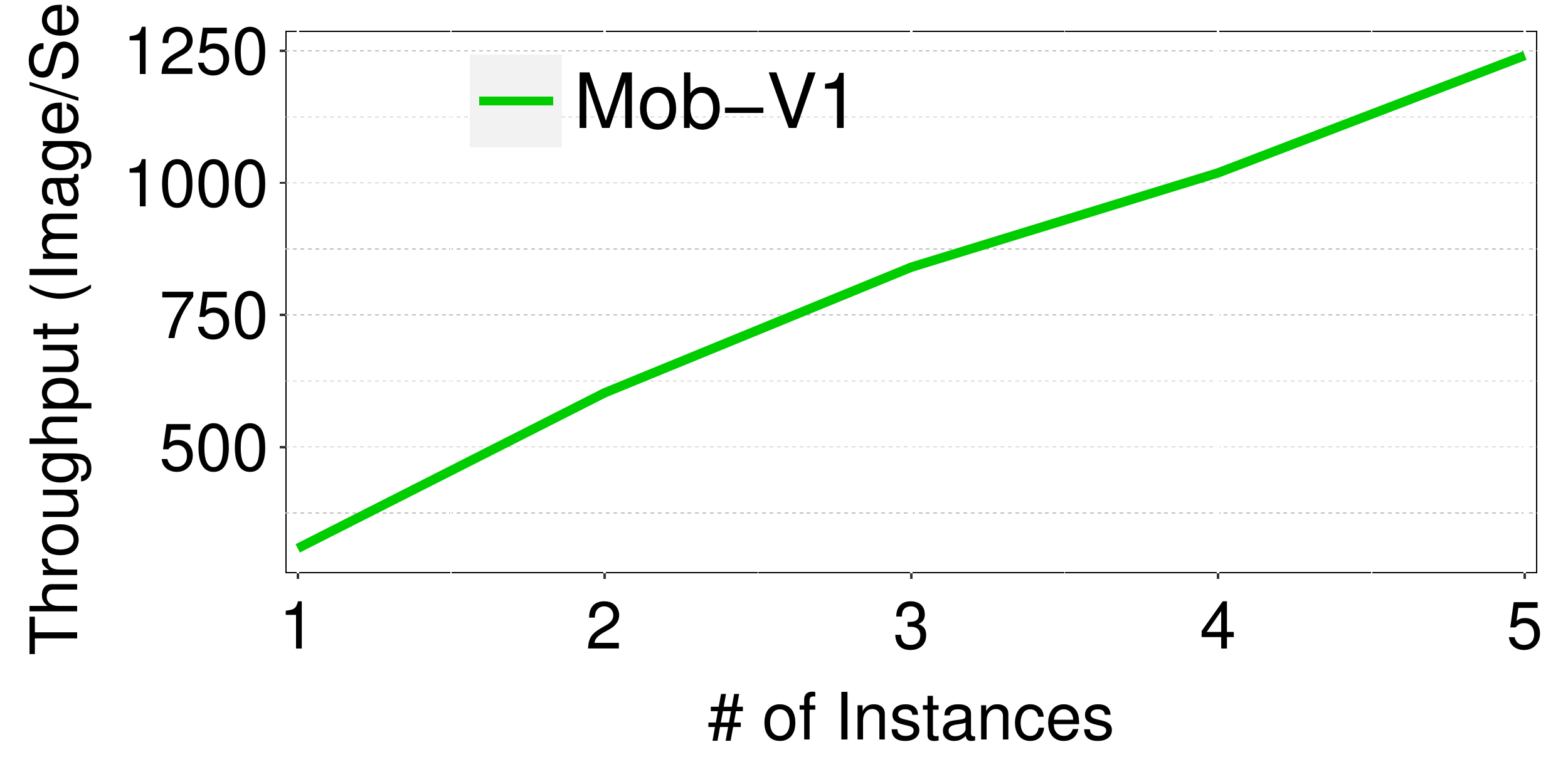}}
    \label{fig:Motivation_Throughput_MTL}}
    \subfloat[Power - Multi-Tenancy]
    {\includegraphics[width=0.45\linewidth]{{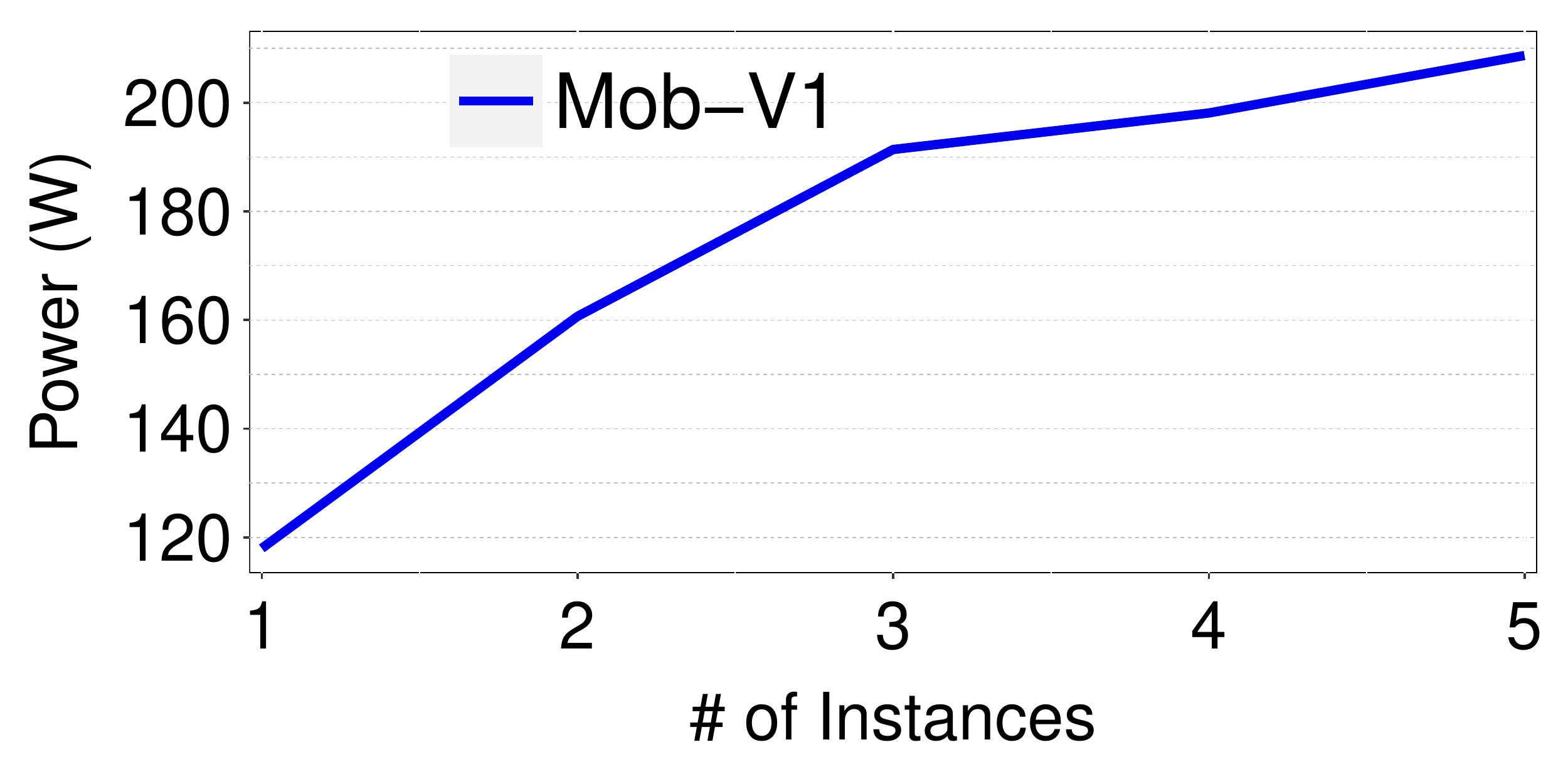}}
    \label{fig:Motivation_Power_MTL}}\\\bigskip
    \subfloat[Throughput - Batching]
    {\includegraphics[width=0.45\linewidth]{{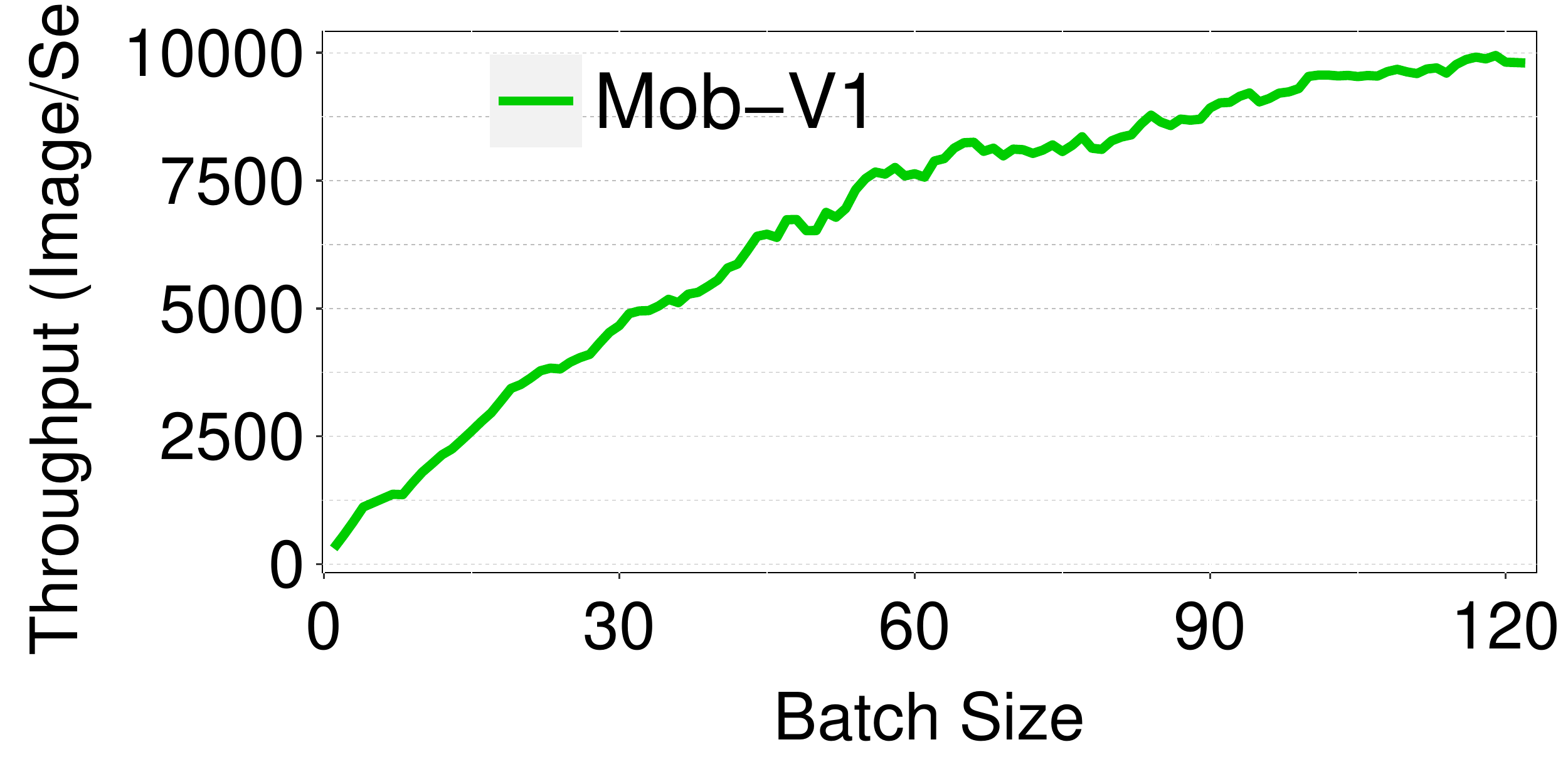}}
    \label{fig:Motivation_Throughput_Batching}}
    \subfloat[Power - Batching]
    {\includegraphics[width=0.45\linewidth]{{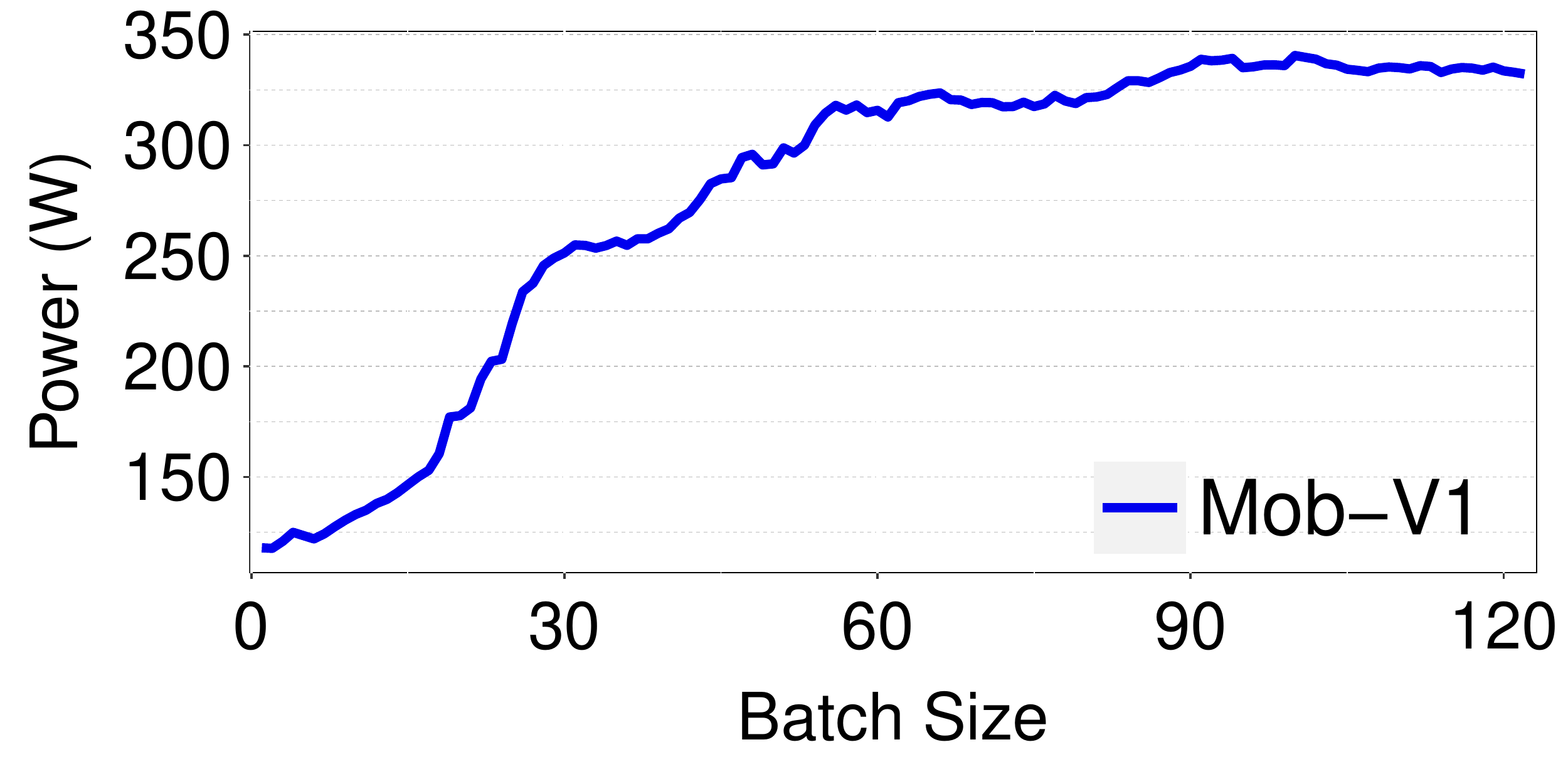}}
    \label{fig:Motivation_Power_Batching}}
\caption{Impact of sole batching and multi-tenancy on throughput and power.}
\label{fig:motivBSMTL}
\end{figure}

\begin{figure}[t]
\centering
    \subfloat[Throughput-MobV1]
    {\includegraphics[width=0.45\linewidth]{{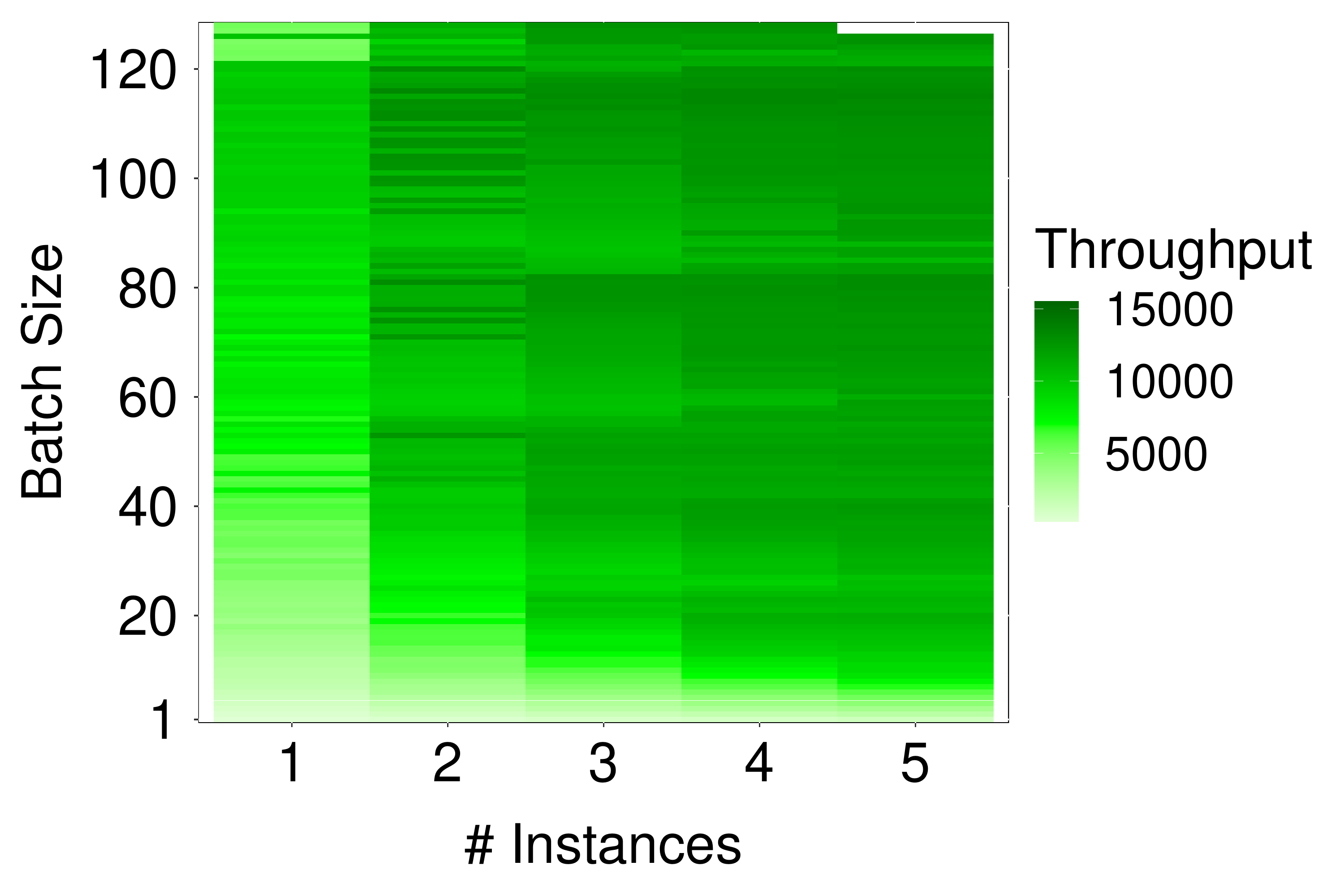}}
    \label{fig:Throughput_Mob_Combine}}
    \subfloat[Power-MobV1]
    {\includegraphics[width=0.45\linewidth]{{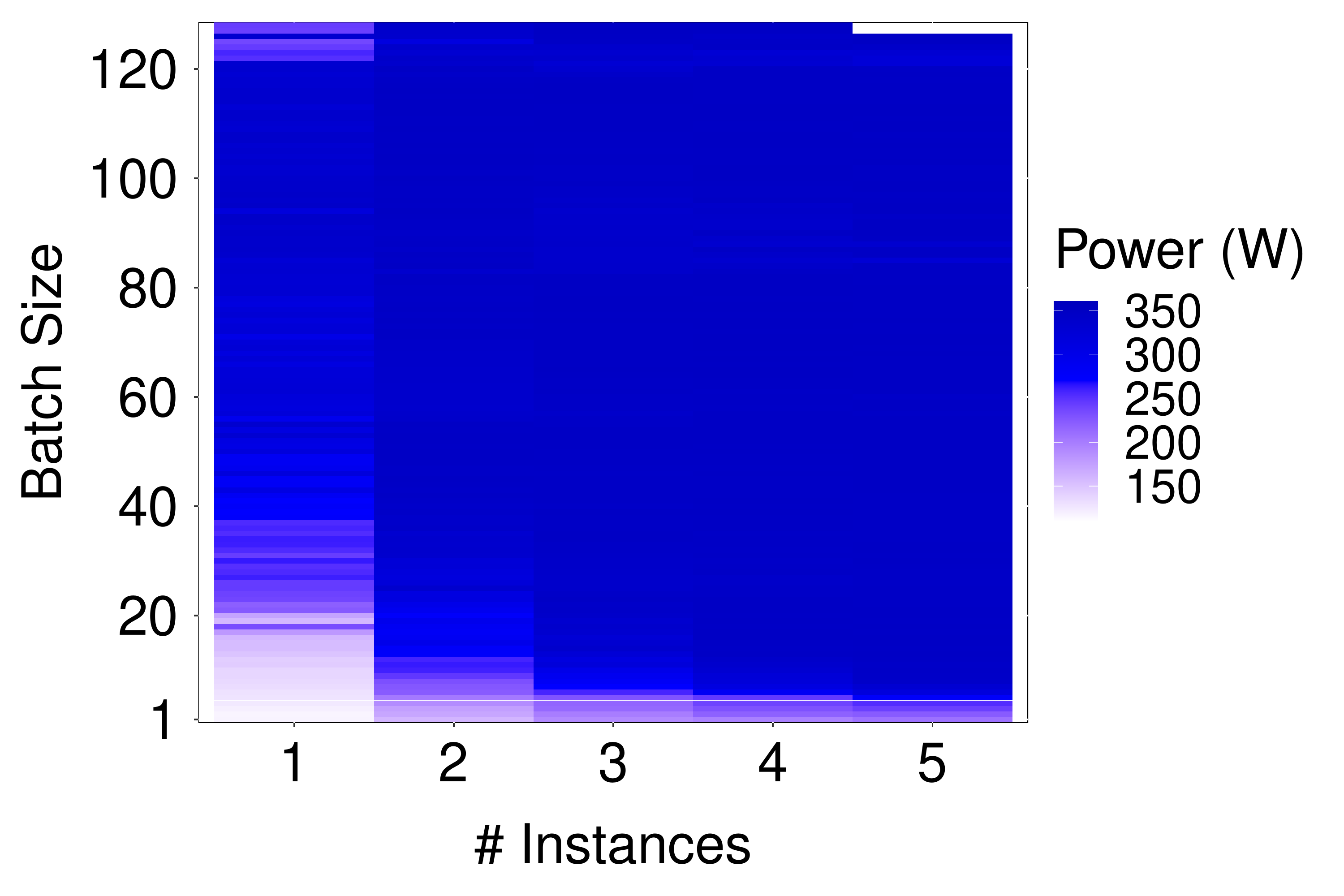}}
    \label{fig:Power_Mob_Combine}}
\caption{Impact of batching and multi-tenancy combination on throughput and power.}
\label{fig:motivComb}
\end{figure}

We conduct a set of preliminary experiments to understand the behavior of these three approaches and their impact on overall throughput. We use the MobileNet-V1 model, an image classification DNN, and measure the maximum throughput it can achieve with each approach. We use images from the ImageNet dataset \cite{russakovsky2015imagenet} and a server equipped with an Nvidia RTX 3090 GPU with 10496 CUDA cores and 24GB GDDR6X memory. Figures~\ref{fig:motivBSMTL} and \ref{fig:motivComb} plot the throughput and power results under different configurations.

While multi-tenancy can increase the throughput, the maximum number of instances that can be deployed on the GPU is limited by the GPU's memory capacity and the models' memory requirement. In our experiments, the maximum number of instances that can be deployed on GPU is five. Therefore, the amount of throughput increase achievable by this approach is limited, and it yields the lowest maximum throughput, 1241 images/sec. Furthermore, we can see that multi-tenancy cannot fully utilize the GPU's maximum power capacity at 350W. Compared with multi-tenancy, batching can almost fully utilize the power capacity of GPU to significantly improve the maximum achievable throughput, up to 9948 images/sec. Finally, we see that combining batching and multi-tenancy can lead to the highest throughput. It achieves a maximum throughput of 13431 images/sec, while its power consumption is the same as the batching approach. Specifically, using both knobs improves the maximum throughput by 982\% and 35\%, compared with multi-tenancy and batching, respectively. These observations emphasize the importance of leveraging both approaches simultaneously to maximize the throughput of DNN inference applications.

\section{Opportunities Under Power Fluctuations}
\label{sec:experiments}

In the previous section, we only showed the maximum throughput each approach could obtain under a specific power limit (350W) to explore their potential to increase the throughput. 
In this section, we go one step further and examine different approaches under fluctuating power caps to see how this potential can lead to higher throughput in the presence of real-world power traces and profiling information. Our trace-driven simulations study the behavior of each approach over one year. In these experiments, the power cap changes every hour as the power generated by each renewable energy source fluctuates. Each DNN is empirically profiled using the \texttt{nvidia-smi} tool on a NIVIDA RTX 3090 GPU to obtain its throughput and power consumption under different batch sizes (BS) and the number of co-located instances (MTL).

\para{Empirical Profiling.}
All the possible combinations of BS and MTL are being considered. A number of instances of the model equal to MTL are created and deployed on the GPU. After that, they have been fed for a specific period of time, e.g., one minute, with batches of input data with a size equal to BS. During the execution, the latency of processing the input batches is measured, and the GPU power consumption is monitored using the GPU power monitoring tool. After the execution is finished, the throughput is calculated with respect to the average latency of processing the input batches and the value of BS and MTL. The power consumption is also calculated by averaging the power consumption monitoring values over the execution time. For example, a model is profiled for batch sizes from 1 to 128 with one instance deployed on GPU, one minute for each batch size, and its throughput and maximum power consumption for that batch size are obtained. The same profiling has been conducted for other configurations, such as two co-located instances and different batch sizes. This profiling data is obtained for every DNN model. Then, each approach uses the data during simulation to determine the power consumption and throughput of each configuration (MTL/BS pair); instead of measuring them while the DNN is being executed on the GPU. The baseline approaches use a part of this profiling data according to their mechanism, e.g., MTL=1 and BS=[1$\cdots$128] for batching.

\para{Power Traces.} 
We use three power traces obtained from ElectricityMap. The duration of the traces is 365 days, and the amount of power capacity is provided for each hour. The power values in traces are normalized to the maximum power consumption of the GPU used in the simulations: 350W for RTX 3090 GPU. Table~\ref{tab:tracesimulation} shows more details of the traces. The variations of traces indicate fluctuation in power capacity, and hence, the power cap applied to the system. We choose these three traces with different variation levels to evaluate the performance of different approaches regarding adaptation to the power cap fluctuation. 

\para{Approaches.} Three approaches are compared in these experiments: multi-tenant, batching, and a combination of them. In all three approaches, an exhaustive search mechanism is embedded that compares the power consumption of all the possible configurations against specified power cap (e.g., all the possible number of instances for the multi-tenant approach, all the possible batch sizes for batching, and all the possible combination of them for combination approach). Then, each approach selects the configurations with the highest throughput and power consumption less than or equal to the power cap. Since all three approaches rely on complete profiling and this profiling imposes significant overhead, they are not applicable in real-world scenarios. However, the results obtained by each approach are the best possible result (highest throughput) that each specific approach can achieve. We compare these upper bound performances of each approach to understand their potential.

\begin{table}[t]
\centering
\caption{Detailed renewable power traces information.}
\label{tab:tracesimulation}
\scriptsize
\begin{tabular}{lrrrr}
\toprule
Trace        & Source       & Mean(W)   & STD   & \begin{tabular}[c]{@{}c@{}}Variation\\ (STD/Mean) * 100\end{tabular} \\ \midrule
US-MIDA-2020 & Hydroeletric & 145.9  & 84.23 & 57.73\%                                                              \\
US-CAL-2020  & Solar        & 239.57 & 76.1  & 31.76\%                                                              \\
SE-2020      & Wind         & 172.92 & 81.96 & 47.39\%                                                              \\ \bottomrule
\end{tabular}
\end{table}

\begin{figure*}[t]
\centering
    \subfloat[Hydroelectric Trace]
    {\includegraphics[width=0.33\linewidth]{{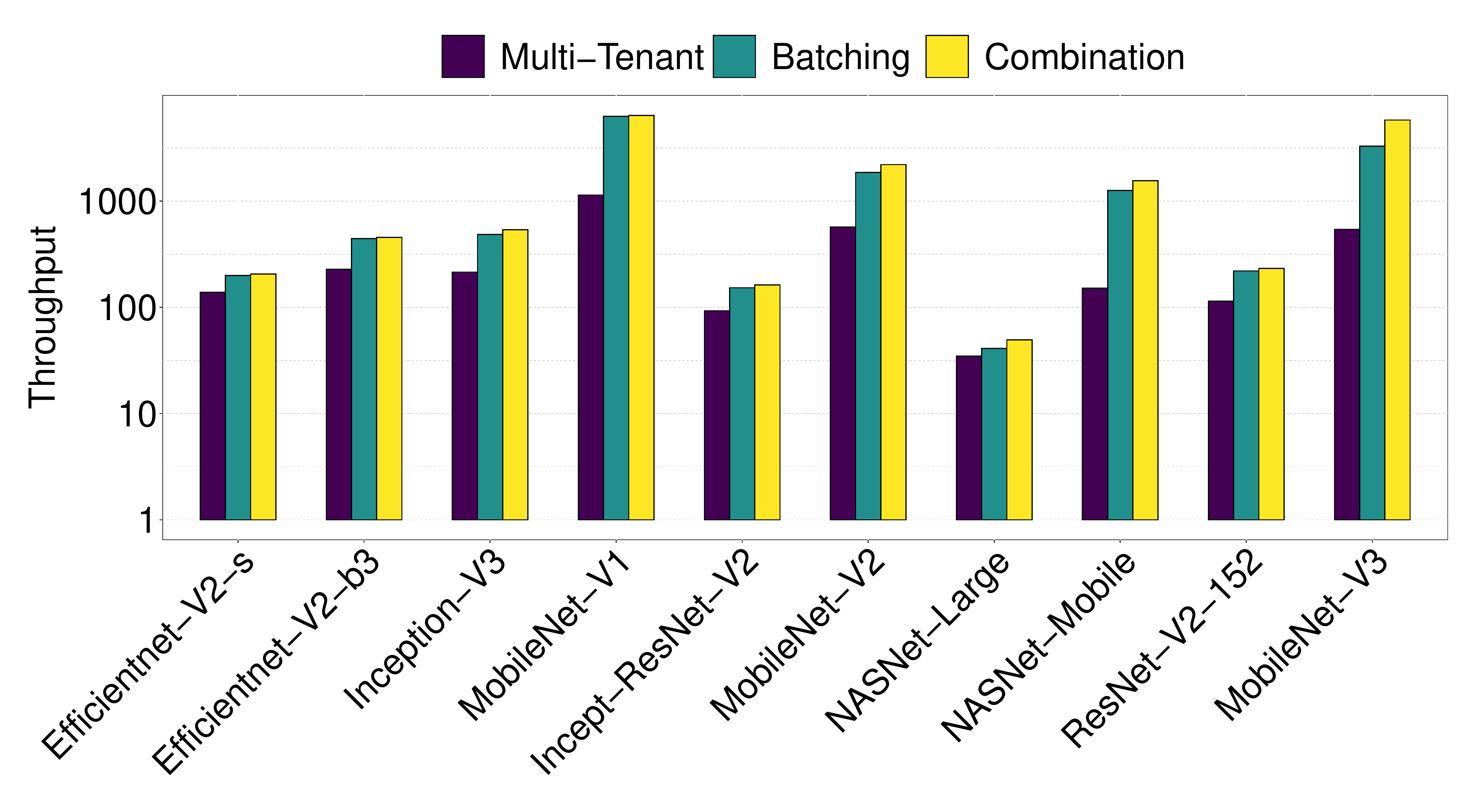}}
    \label{fig:simulationHydro3090}}
    \subfloat[Solar Trace]
    {\includegraphics[width=0.33\linewidth]{{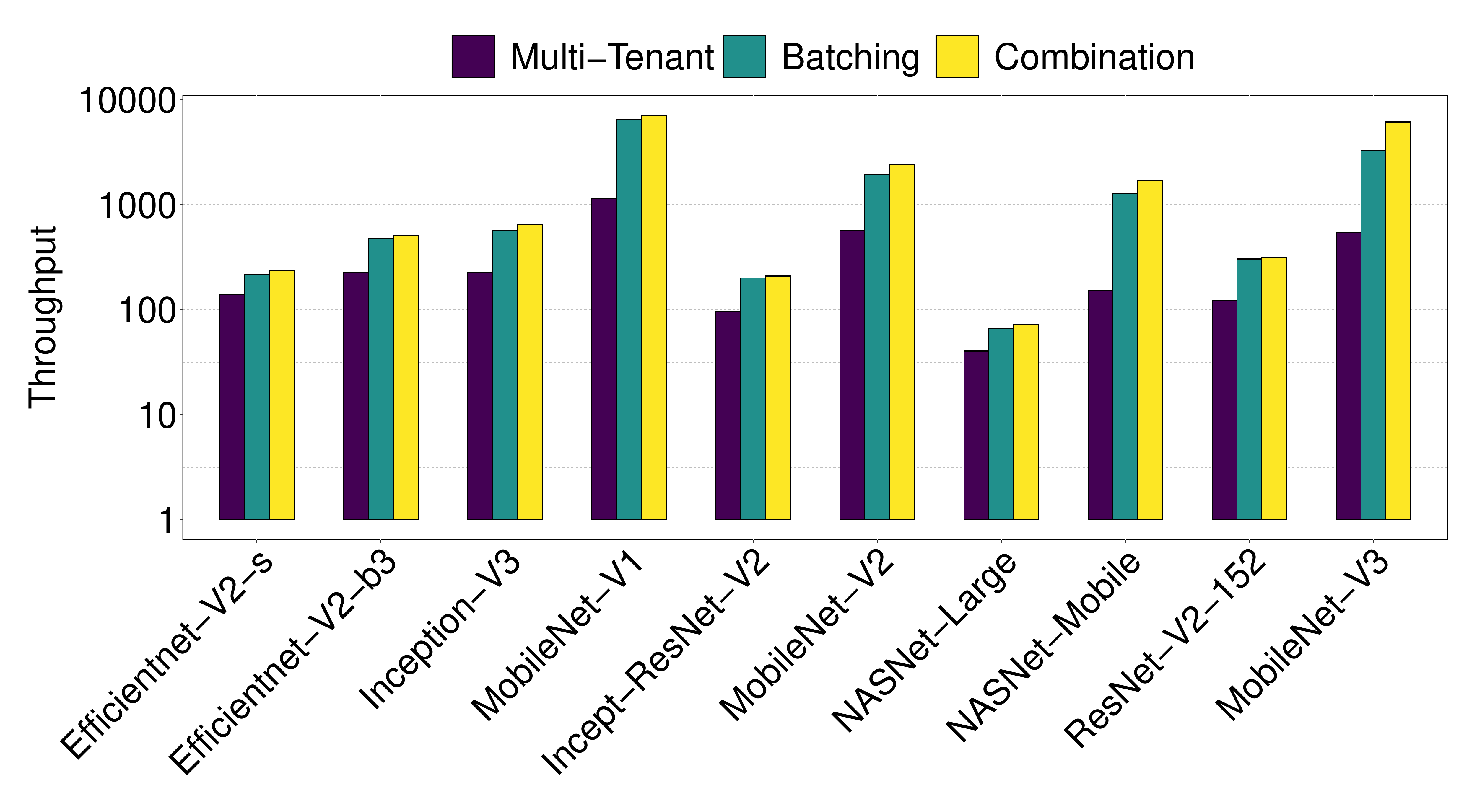}}
    \label{fig:simulationSolar3090}}
    \subfloat[Wind Trace]
    {\includegraphics[width=0.33\linewidth]{{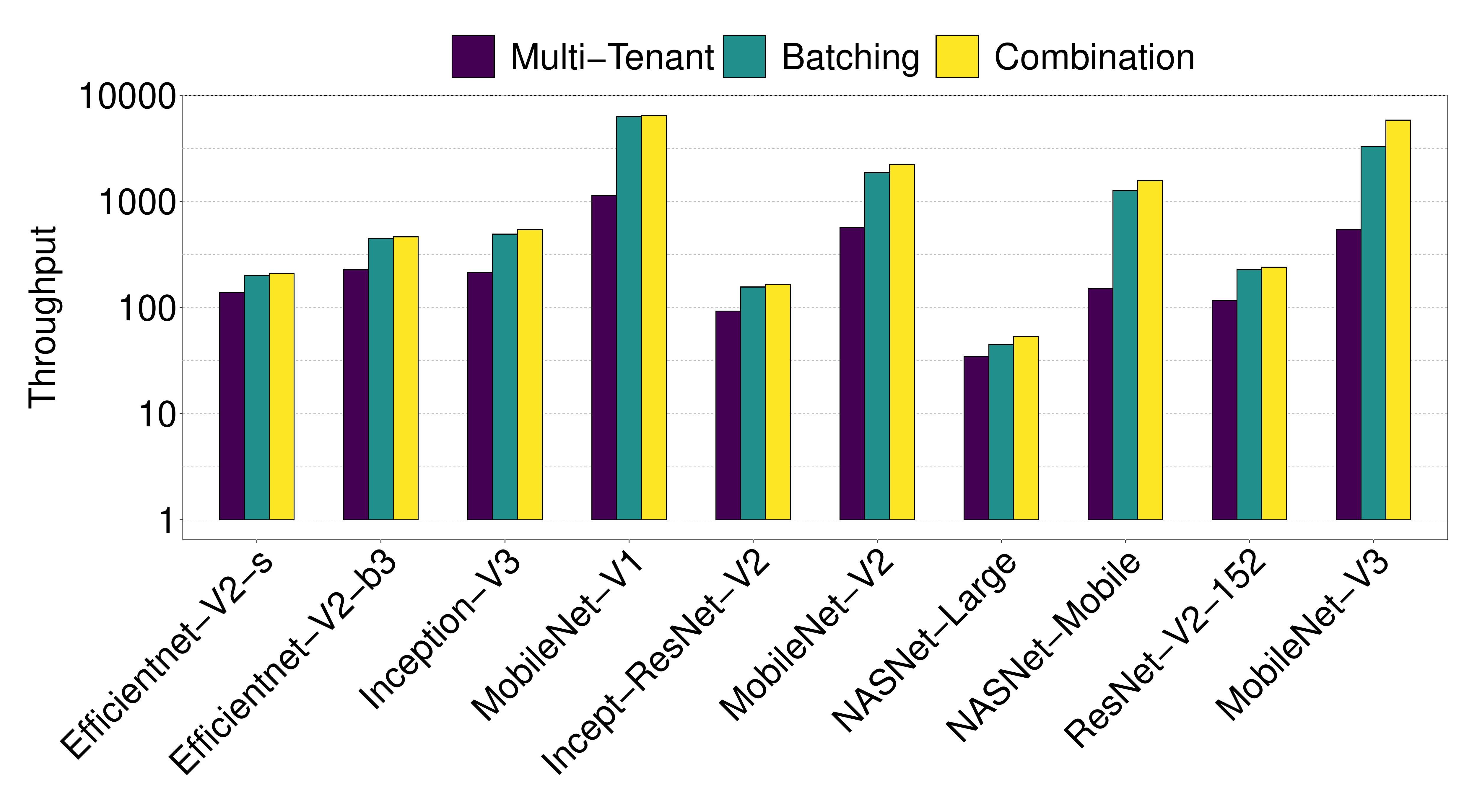}}
    \label{fig:simulationWind3090}}
\caption{Average throughput achieved in large-scale simulation experiments by each approach for RTX 3090 GPU. The Y-axis is shown in the base-10 log scale.} \label{fig:simulationThroughput3090}
\end{figure*}

\para{Results.}
The average throughput (over the entire trace duration) of each DNN is shown in Figure~\ref{fig:simulationThroughput3090}. The average improvement of Combination compared with Batching is 16\%, 19\%, and 17\% for Hydroelectric, Solar, and Wind traces, respectively. The comparison with Multi-Tenant also shows 316\% improvement in Hydroelectric, 363\% in Solar, and 321\% in Wind trace. Considering each individual model, we see that some can benefit more than others from Combination. For example, Combination can improve the throughput by 86\% in Solar trace for MobileNet-V3 compared with Batching. However, in the same trace, it can only improve the throughput by 8\% for NASNet-Large. These results suggest that the specifications of DNN models, such as memory requirement and computational complexity, affect the significance of the benefit they can gain from the combination of batching and multi-tenancy. 

Another observation from Figure~\ref{fig:simulationThroughput3090} is the impact of the significance of the power cap on throughput improvement achievable by combining batching and multi-tenancy. The throughput improvement of Combination, compared with two other approaches, in the solar trace is more than the two other traces. Looking at Table~\ref{tab:tracesimulation}, we see that the mean of power cap in solar trace (239.57W) is more than hydroelectric (145.9W) and wind (172.92W). This higher power cap means more possible combinations of batching and multi-tenancy, providing a larger state space for the Combination approach to explore. Hence, it can find solutions with higher throughput than the others. On the other hand, a tight power cap leads to limited space for the combination of batching and multi-tenancy to achieve high throughput.

\begin{figure}[t]
\centering
    \subfloat[Throughput Results]
    {\includegraphics[width=0.9\linewidth]{{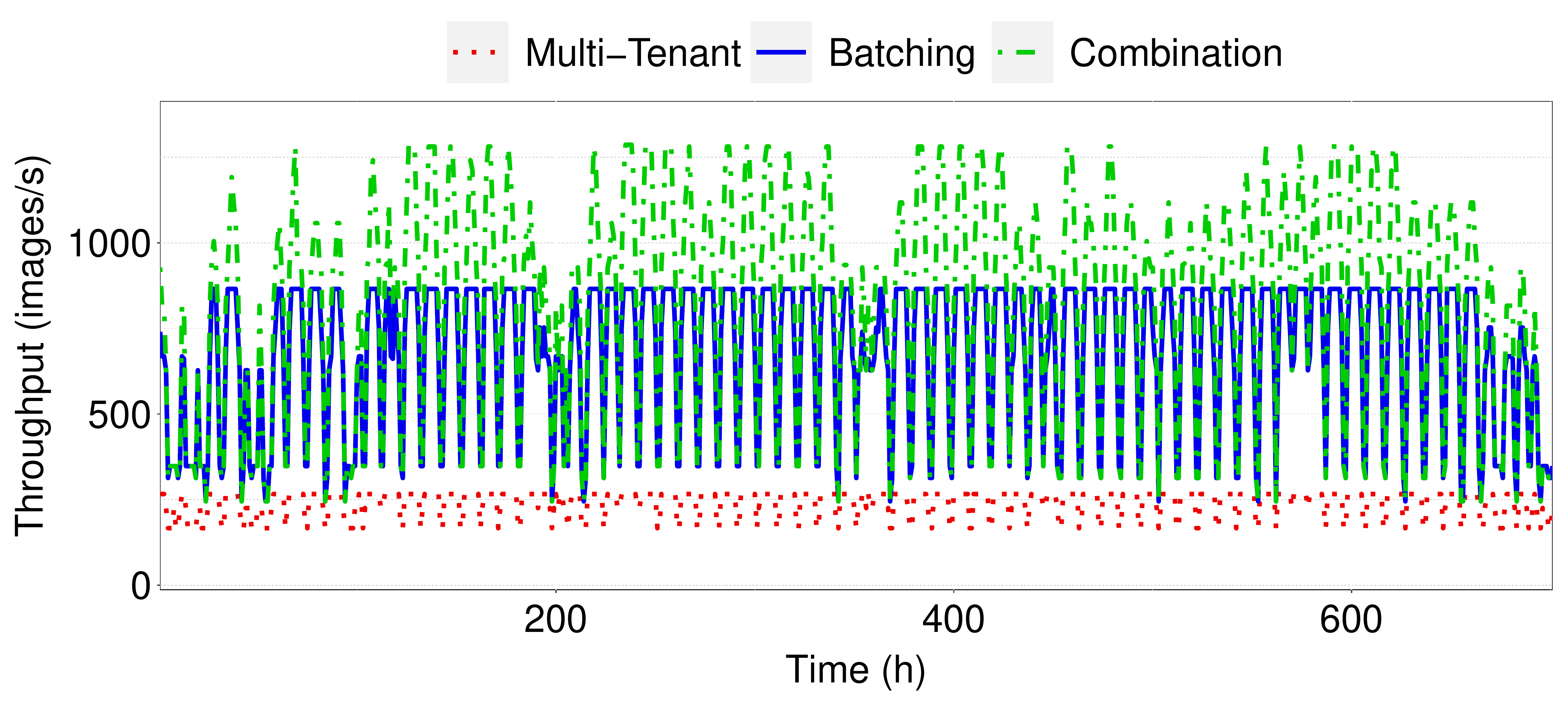}}
    \label{fig:detailthrou}}\\\bigskip
    \subfloat[Power Cap Pattern]
    {\includegraphics[width=0.9\linewidth]{{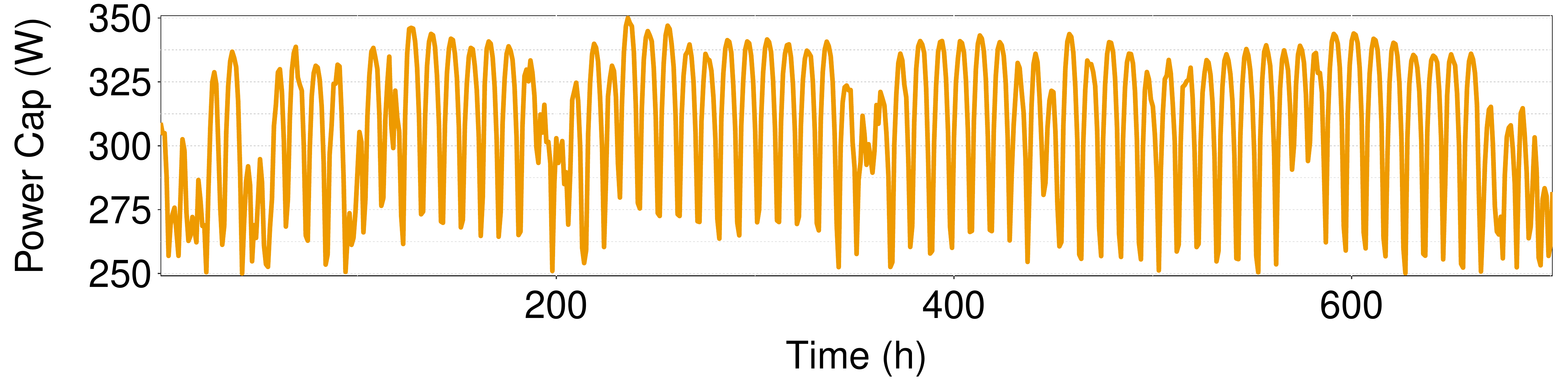}}
    \label{fig:detailpower}}\\
\caption{Detailed throughput achieved by each approach for Inception-V3 model over a time slice of solar trace. The corresponding power cap specified in the trace for that period is also depicted below the throughput results.} \label{fig:detail}
\end{figure}

Finally, we study the detailed throughput of different approaches to understand how the fluctuation of the power cap affects the throughput obtained by each approach over time. To this end, we show the throughput of all three approaches for the Inception-V3 model during a time slice of solar trace in Figure~\ref{fig:detail}, along with the value of the power cap during that time slice. There is a significant fluctuation in throughput for all three approaches. However, the average throughput of the Combination is more than the two other approaches. When the power cap is low, see Figure~\ref{fig:detailpower}, the throughput obtained by Combination is very close to the Batching. However, when the power cap is more relaxed, we see that the Combination approach can achieve throughput significantly higher than Batching. For those power caps, the maximum throughput achievable by Batching is limited to a certain number, and large batch sizes cannot help to further increase it. However, combining batching and multi-tenancy provides the opportunity to obtain higher throughput. This observation emphasizes the importance of leveraging the combination of batching and multi-tenancy to maximize the throughput while meeting the power cap.

\section{Future Research Directions} 
\label{subsec:future}

Based on the observations from \S\ref{sec:experiments}, we propose the following research directions for future work.

Currently, we use power and throughput information of configurations obtained through complete profiling. In the complete profiling, the DNN model is deployed on the GPU, and all the possible combinations of MTL and BS are sampled and executed to collect their throughput and power consumption. Having access to complete profiling of configurations can lead to significantly higher throughput. However, the time and resource overhead of such profiling might not be acceptable for all workloads. This profiling cost can be amortized over time for recurrent long-running jobs that use the same DNN. However, for shorter jobs with a new DNN model that has not been profiled before, the overhead imposed by profiling all the possible configurations might exceed the job's runtime. Furthermore, the profiling space for MTL and BS can be larger for more powerful GPUs and smaller models, leading to significantly higher overhead. For comparison, it takes around 9 hours to profile a model with a maximum MTL of 4 and BS of 128 on the 3090 GPU.  
Therefore, a more sophisticated approach is needed to explore the configurations' state space online and with low overhead. Any proposed approach should be able to traverse the MTL and BS simultaneously to ensure that the entire state space can be explored during the search period. Finding a value for each control knob one by one might end up in a local optimal.

\para{Reducing Profiling Overhead.}
One possible approach to address this challenge is using sampling-based approaches. The approach selects several configurations as samples and tests them to find a solution that maximizes the throughput while meeting the power cap. Different sample selection methods can be designed using heuristic algorithms or mathematical models. For example, it is possible to define a new metric based on the relationship between control knobs (MTL and BS) and the throughput and power consumption. Then, the configurations state space can be explored based on this metric to select the sample configurations. Other approaches, such as ones based on Bayesian Optimization, can be designed to select the sample configurations more methodically based on their proximity to the possible optimal or near-optimal solution.

\para{Desigining Performance Models.}
Another possible approach is leveraging ML-based approaches to predict different configurations' throughput and power consumption. This approach helps to reduce the time and resource overhead of testing sample configurations imposed by profiling-based approaches. When designing these ML models, we can consider different features, such as the batch size, the number of instances, the number of parameters of the network, the number and type of layers, its computing and memory utilization, and even low-level hardware metrics, such as the pattern of access to the cache. Using DNNs for prediction is an option. However, for the predictions to be accurate enough, we need to gather a large amount of training data to obtain sophisticated models with high accuracy. The process of gathering the training data can be challenging as the required dataset should be large enough to represent various features and their effect on power consumption and throughput.

\para{Handling Power Cap Violations.}
The other challenge is power cap violation which happens when the power consumption goes above the power cap indicated for the system due to excessive usage of resources under the configuration. The power cap violation problem can happen in two cases: \1 the power cap determined by an external controller changes and comes below the current power consumption of the system. For example, the amount of power generated by wind turbines can change during different hours of the day, which affects the power cap allocated to the system. \2 Due to some internal hardware (e.g., increased GPU temperature) or application (change in the source of input data) changes, the power consumption exceeds the power cap. To address this challenge, two different approaches can be taken into account: \1 Proactive: in this approach, both the power consumption and power cap are constantly monitored by a monitoring unit over time. The monitoring unit predicts future power cap violations based on past data and warns the configuration selection unit to readjust the configuration to avoid the violation. \2 Reactive: The power monitoring unit only sends a signal to the configuration selection unit after a power cap violation happens and is detected so that the unit would readjust the configuration to stop the power cap violation.

\section{Related Work}\label{sec:related}

A large body of research has focused on handling the intrinsic instability of renewable energy to support cloud and datacenter \cite{gao2020smartly, kwon2020ensuring, he2022online, chakraborty2023elastic, souza2023ecovisor, kwon2018demand, souza2023ecovisor}. To address this challenge, Gao et al.\cite{gao2020smartly} propose the joint consolidation of jobs with the same SLO to the same physical machine (PM) and the assignment of renewable energy generators to the PMs. It consolidates the jobs with a certain SLO and predicted energy demand on the same PM. Then, It assigns the renewable energy generator that is predicted to yield no less than the predicted energy demand over time. 
Unlike this work, Elastic Power Utilization (EPU) approach \cite{chakraborty2023elastic} tries to manage the workload of data center, so the energy demand of workloads would be similar to the energy available from renewable sources (instead of matching supplies with PMs, it matches PMs with supplies). 
Kwon et al. \cite{kwon2020ensuring} targets the improvement of renewable energy utilization in data centers by integrating solar power and batteries. To this end, it proposes a mathematical optimization model to minimize energy costs while guaranteeing the desired level of renewable energy utilization and the required quality of service (QoS) guarantee. 
All these approaches leverage system-level techniques such as consolidation, resource provisioning, or workload scheduling to address the fluctuation of renewable energy. However, in our proposed approach, we leverage the control knobs at the application level to manage its power demand with respect to power source limitations. 

To improve the performance of DNN inference on GPU accelerators, prior work has explored techniques such as batching and multi-tenancy \cite{gujarati2020serving, nabavinejad2022coordinated, li2022miso, gupta2020deeprecsys, crankshaw2017clipper, choi2022serving, wang2021morphling}. Morphling \cite{wang2021morphling} introduces an approach for configuration tuning of DNN inference systems that aims for fast, near-optimal auto-configuration. It converts the problem of finding optimal configuration to a few-shot learning problem and then solves it using a model-agnostic meta-learning technique. 
iGniter \cite{xu2022igniter} is an interference-aware GPU resource provisioning framework that aims to make the performance of DNN inference workloads predictable while minimizing the resource usage of GPU accelerators, which leverages batching. BatchDVFS \cite{nabavinejad2022coordinated} leverages the combination of batching and dynamic voltage frequency scaling (DVFS) to manage the power consumption of DNN inference applications below a certain power cap, while maximizing the throughput. 
MISO \cite{li2022miso} explores a new partitioning capability of Nvidia GPUs (e.g., A100) called Multi-Instance GPU (MIG). MISO proposes an ML-based mechanism to predict the performance of an application on a MIG partition. Using the prediction results, it dynamically deploys applications on GPU partitions to improve performance. Unlike these approaches that consider either batching or multi-tenancy, our proposed solution considers the combination of them. Even the works that leverage multi-tenancy, consider the co-location of several different models on the same GPU, but we consider several instances of the same model. Finally, no previous work leverages the combination of batching and multi-tenancy to control the power consumption with respect to a power cap, while maximizing the throughput.

\section{Conclusion}\label{sec:conc}

We investigated the impact of joint batching and multi-tenancy on the performance of DNN inference applications under a fluctuating power cap. The results show that combining these two control knobs provide more room to increase the throughput while meeting the power constraint. The combination approach allows us to better utilize the GPU resources and available power capacity. We also shed light on some of the challenges that needed to be addressed to better utilize the combination approach. These include how to efficiently explore the configurations' state space and find a proper solution with a few samples or using ML-based prediction without sampling, and how to predict or detect the power cap violations in real-time and prevent or address them.

\begin{acks}
We thank ElectricityMap for their carbon intensity dataset. This work is partly supported by NSF Grants 2105564, 2236987, and VMWare
\end{acks}

\bibliographystyle{ACM-Reference-Format}
\bibliography{sample-base}

\end{document}